\documentclass{jpsj3}
\usepackage{txfonts}
\usepackage{color}
\usepackage{braket}
\usepackage{bm}

\title{Observations of $\nu=1$ Quantum Hall Effect and Inter-Band Effects of Magnetic fields on Hall Conductivity in Organic Massless Dirac Fermion System $\alpha$-(BETS)$_2$I$_3$ under Pressure}

\author{Kazuyuki Iwata$^1$, Akito Koshiba$^1$, Yoshitaka Kawasugi$^1$, Reizo Kato$^2$, and Naoya Tajima$^1$\thanks{naoya.tajima@sci.toho-u.ac.jp}}

\inst{
$^1$Department of Physics, Toho University, Funabashi, Chiba 274-8510, Japan.\\ 
$^2$RIKEN, Hirosawa 2-1, Wako-shi, Saitama 351-0198, Japan.
}

\abst{
We investigated the magnetoresistance and the Hall effect in an organic massless Dirac fermion system $\alpha$-(BETS)$_2$I$_3$ under pressure. 
The Fermi energy of this system is slightly far away from the Dirac points, and thus the $\nu =1$ quantum Hall state is realized in a low magnetic field at low temperatures. 
Moreover, the experimental formula for chemical potential as a function of temperature is clarified.
We succeeded in detecting the inter-band effects of the magnetic field on the Hall conductivity when the chemical potential passes the Dirac points.
}

\allowdisplaybreaks

\newcommand{\be}{ \begin{equation}}
\newcommand{\ee}{ \end{equation}}

\usepackage{color}

\usepackage{ulem}

\begin{document}
\maketitle

The discovery of graphene\cite{rf:1, rf:2} opened up a new field of "massless Dirac fermion" in condensed matter physics. 
A wide variety of materials with massless Dirac fermions have been now established. 

The massless Dirac fermion system has a unique energy structure with two cone-type bands touching each other at discrete Dirac points in the first Brillouin zone. 
The low-energy excitations described by a relativistic Dirac equation give rise to the $\pi$ Berry phase that characterizes the transport of the Dirac fermions. 

In such a system, we are interested in how the Dirac point becomes unstable and creates a mass gap, e.g., through electron-electron interactions or spin-orbit coupling.

To address this question, we should target Dirac materials in which the electron-electron interaction can be controlled and the Fermi energy is in the Dirac points. 
To our knowledge, only organic massless Dirac fermion systems can be responded those requests.

Among them, $\alpha$-(BEDT-TTF)$_2$I$_3$\cite{rf:3} undergoes a charge-order phase transition at 135 K under ambient pressure\cite{rf:4, rf:5, rf:6, rf:7, rf:8, rf:9}, and a massless Dirac fermion system is realized at pressures above 1.5 GPa\cite{rf:10, rf:11, rf:12, rf:13, rf:14, rf:15, rf:16, rf:17, rf:18, rf:19}.
This Dirac fermion system undergoes a crossover from a two-dimensional (2D) Dirac semimetal to a 3D Dirac semimetal at low temperatures\cite{rf:20, rf:21}.
This is the first material in which the Dirac fermion phase is next to the strongly correlated electron phase, as in the charge-ordered insulator phase.
The physics of strongly correlated Dirac fermions is unfolding in this material.
For example, NMR experiments and theory suggest that the excitonic phase is realized at low temperatures and under magnetic fields\cite{rf:22, rf:23, rf:24}.

However, this system gave us an unexpected answer to the above question.
When the massless Dirac fermion phase in this system approaches the charge-ordered insulating phase by releasing pressure, the Fermi velocity decreases without creating the mass gap\cite{rf:25}.
The decrease of the Fermi velocity is associated with the strong on-site Coulomb repulsion\cite{rf:26}.

In contrast, it has been suggested that the sister compound $\alpha$-(BETS)$_2$I$_3$ [BETS = Bis(ethylenedithio)tetraselenafulvalene
]\cite{rf:27} under the ambient pressure was the Dirac fermion system with a mass gap of approximately 50 K\cite{rf:28, rf:29, rf:30}. 
So, this material becomes insulating at temperatures below 50 K.  
The discovery of electromagnetic duality experimentally and theoretically that is valid only within the relativistic framework strongly suggests that a massive Dirac fermion system is realized in this material at a temperature below 50 K\cite{rf:30}.
According to the first principles calculations, spin-orbit coupling in this system plays an important role in the creation of a mass gap\cite{rf:28, rf:29, rf:31}. 
However, Ohki et al. indicated theoretically that the correlation effects also play an important role in mass gap creation\cite{rf:32, rf:33}. 
At temperatures below 50 K, the detection of magnetic orders by Konoike et al. would suggest that electron correlations play an important role in the mass gap creation in this system\cite{rf:34}.

Under pressures above 0.5 GPa, the insulating behavior of this material is suppressed and a massless Dirac fermion system is realized.\cite{rf:12, rf:35, rf:36, rf:37}.
Detecting the effects of $n=0$ Landau level called zero modes on the in-plane transport is conclusive evidence\cite{rf:36}. 
The phase of the Dirac fermions is confirmed from the Shubnikov-de Haas (SdH) oscillation of the carrier-doped samples\cite{rf:37}. 

Thus, by controlling electron correlation effects and/or spin-orbit coupling with pressure, this material provides a testing ground for experimental investigations of how the Dirac point becomes unstable and creates a mass gap.

For this purpose, it is necessary to obtain detailed information such as how far the Fermi energy is from the Dirac point under high pressure.
It will also be important to observe quantum transport phenomena in this situation.

In this letter, we experimentally demonstrate that the Fermi energy of this system under high pressure is located approximately 1.4 K from the Dirac points. Thus the $\nu =1$ quantum Hall effect is realized at low temperatures.
In general, the quantum Hall effect is inherent to two-dimensional electron systems.
It is very rare to observe the quantum Hall effect in bulk crystals.
This system has a layered structure where each layer is independent and thus has a strong two-dimensional nature. It allowed us to observe the quantum Hall effect. 
Moreover, the temperature dependence of the chemical potential is revealed. 
We succeeded in detecting the inter-band effects of the magnetic field on the Hall conductivity when the chemical potential passes the Dirac points.

A sample on which electrical leads were attached was put in a Teflon capsule filled with the pressure medium (Idemitsu DN-oil 7373) and then the capsule was set in a clamp-type pressure cell made of hard alloy MP35N.

We measured the resistance and the Hall resistance of three samples under the pressure of 1.7 GPa by a conventional DC method with six probes. 
An electrical current of 10 $\mu$A was applied along the $a$-axis and the magnetic field was applied along the $c$-crystal axis normal to the plane.  

\begin{figure}[htbp]
  \includegraphics[width=0.8 \linewidth]{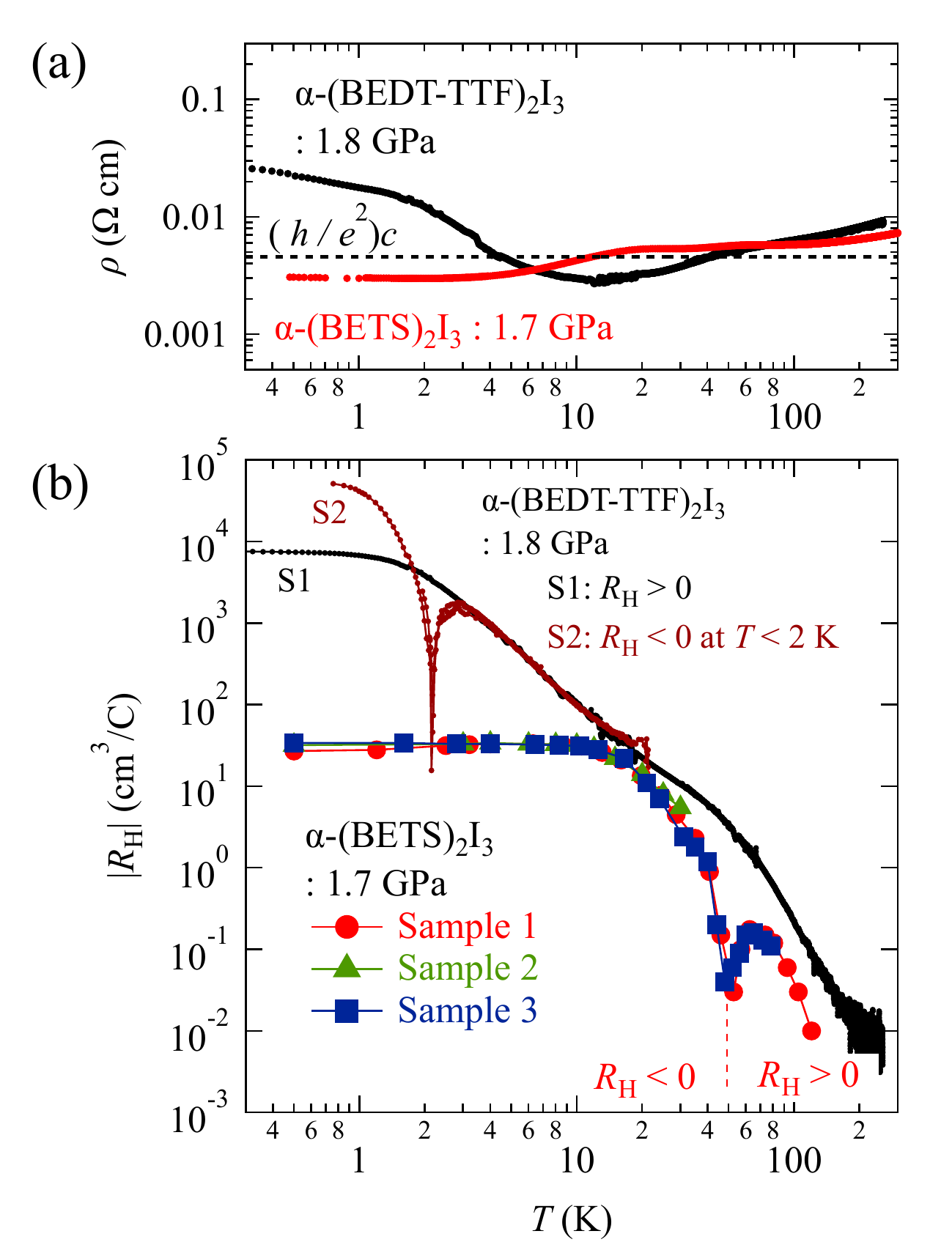}
  \caption{
    \label{fig:1}
    (color online) (a) Temperature depemdence of the resistivity and (b) the Hall coefficient $R_H$ of $\alpha$-(BEDT-TTF)$_2$I$_3$ for $p=$ 1.8 GPa and $\alpha$-(BETS)$_2$I$_3$ for $p=$ 1.7 GPa. 
    $\alpha$-(BEDT-TTF)$_2$I$_3$ samples are classified as hole doped (S1) or electron doped (S2) types depending on the dopant due to the instability of the I$_3$$^-$ anion. 
    Note that the absolute value of $R_H$ is plotted. 
    The dips indicate a change in the polarity. 
    In S2, the polarity is changed at approximately 2 K.
    All $\alpha$-(BETS)$_2$I$_3$ samples, on the other hand, change their polarity at approximately 50 K.
  }
\end{figure}

Figure 1 shows (a) the temperature dependence of the resistivity and (b) the Hall coefficient $R_H$ of $\alpha$-(BEDT-TTF)$_2$I$_3$ for $p=$ 1.8 GPa and $\alpha$-(BETS)$_2$I$_3$ for $p=$ 1.7 GPa. 
The $R_H$ of the two $\alpha$-(BEDT-TTF)$_2$I$_3$ samples S1 and S2 are data in Ref. \cite{rf:38}. 
On the other hand, the $R_H$ of the three $\alpha$-(BETS)$_2$I$_3$ samples were examined from the linear Hall resistivity at around zero-field. 

We see that the two-crystal resistivity depends on the temperature very weakly. The value is close to $(h/e^2)c$, where $h/e^2 = 25.8$ k$\Omega$ is the quantum resistance and the $c=1.78$ nm is a lattice constant for $c$-axis. This is a characteristic transport in 2D massless Dirac fermion systems with a layered structure where the Fermi energy, $E_F$, is at or near the Dirac points ($E_F \sim 0$)\cite{rf:39}.

Note that in $\alpha$-(BEDT-TTF)$_2$I$_3$, the I$_3$$^-$ anion is unstable, resulting in a strong sample dependence of the resistivity at low temperatures\cite{rf:38}. 
However, the resistivity in $\alpha$-(BETS)$_2$I$_3$ shows almost no sample dependence. 
This suggests that the $E_F$ of $\alpha$-(BETS)$_2$I$_3$ is away from the Dirac point than that of $\alpha$-(BEDT-TTF)$_2$I$_3$, and the density of states there is expected to be so high that ppm-level doping has no effect on the resistivity.
In the following, we investigate the $E_F$ of $\alpha$-(BETS)$_2$I$_3$ from the Hall effect.

Because it should be $R_H=0$ at $E_F=0$, the high Hall coefficients at low temperatures suggest $E_F \neq 0$. 
The $R_H$ of $\alpha$-(BETS)$_2$I$_3$ is more than two orders of magnitude lower than that of $\alpha$-(BEDT-TTF)$_2$I$_3$, which suggests that the $E_F$ of $\alpha$-(BETS)$_2$I$_3$ is much higher than that of $\alpha$-(BEDT-TTF)$_2$I$_3$. 

Now, we examine $E_F$ for $\alpha$-(BETS)$_2$I$_3$ in a 2D Dirac Fermion regime as follows. 
The strong temperature dependence of the $R_H$ shown in Fig. 2(b) will be discussed later.

The first step is to estimate the sheet density, $n_s$, from $R_H$ as $n_s=(1/R_H e)c$. 
A conductive layer of this material is sandwiched by insulating layers and therefore, each conductive layer is almost independent. 
The ratio of in-plane to out-of-plane conductivity is several thousand times greater. 
So, sheet density (density per layer) is valid for this system.

Secondly, the area of the Fermi surface $S_F=n_s/4$ and the Fermi wave number $k_F=\sqrt{S_F/\pi}$ are obtained, where factor 4 indicates the degeneracy of spins and valleys. 

Finally, we estimate $E_F=\hbar v_F k_F$ with the Fermi velocity $v_F=3.6 \times 10^4$ m/s. 
We show $R_H$, $n_s$, $S_F$, $k_F$, $E_F/k_B$ for three samples at 0.5 K in Table 1. 

%
\begin{table}[h]
 \caption{$R_H$, $n_s$, $S_F$, $k_F$ and $E_F$ for three samples at 0.5 K}
 \label{table:ef}
 \centering
  \begin{tabular}{cllll}
   \hline
  quantities & sample 1 & sample 2 & sample 3 & units \\
   \hline \hline
   $R_H$ & 27 & 32 & 34 & cm$^3$/C \\
   $n_s$ & 4.1 & 3.5 & 3.3 & $\times 10^{10}$ cm$^{-2}$ \\
   $S_F$ & 1.0 & 0.88 & 0.83 & $\times 10^{10}$ cm$^{-2}$ \\
   $k_F$ & 5.6 & 5.3 & 5.1 & $\times 10^{6}$ m$^{-1}$ \\
   $E_F/k_B$ & 1.5 & 1.4 & 1.3 & K \\
   \hline
  \end{tabular}
\end{table}
%

The average $E_F/k_B$ of the three samples is 1.4 K. 
This result motivates us to observe the quantum Hall effect at low-filling factors in bulk crystals.

In the massless Dirac fermion systems, the appearance of an $n=0$ Landau level at the Dirac point indicates a special electron-hole degenerate Landau level due to the exceptional topology of the linear band structure. 
On the other hand, quantum Hall states near the Dirac point are affected by the strong electron correlation.
Hence, the realization of the quantum Hall states corresponding to a resolving of the spin and valley symmetry degenerate $n=0$ Landau levels are expected in this system. 

Figure 2 shows the polarized $n=0$ Landau levels of spins ($\uparrow$ and $\downarrow$) and valleys ($+$ and $-$), as expected from a detailed analysis of SdH oscillations in the carrier-doped sample\cite{rf:37}.
Spins polarized energy is expressed as $\Delta_z=2\mu_B B$, where $\mu_B$ is the Bohr magneton. 
On the other hand, the Coulomb interaction in the magnetic field yields valley polarization, whose polarization energy, $\Delta_v$, depends on the square root of the magnetic field as $\Delta_v/k_B=a\sqrt{B}$\cite{rf:40}, where $a$ is estimated to be $\sim 0.64$ KT$^{-1/2}$ from an analysis of SdH oscillations in the carrier-doped sample\cite{rf:37}.
The polarized $n=0$ Landau levels shown in Fig. 2, $n_{0\downarrow \pm}$ and $n_{0\uparrow \pm}$, are written as $E_{0\downarrow \pm}=\Delta_z/2 \pm \Delta_v/2$ and $E_{0\uparrow \pm}=-\Delta_z/2 \pm \Delta_v/2$.

In the following, we experimentally demonstrate that the $\nu =1$ quantum Hall state prospected in Fig. 2 is realized in a low magnetic field at a temperature below 1 K.

\begin{figure}[htbp]
  \includegraphics[width=0.7 \linewidth]{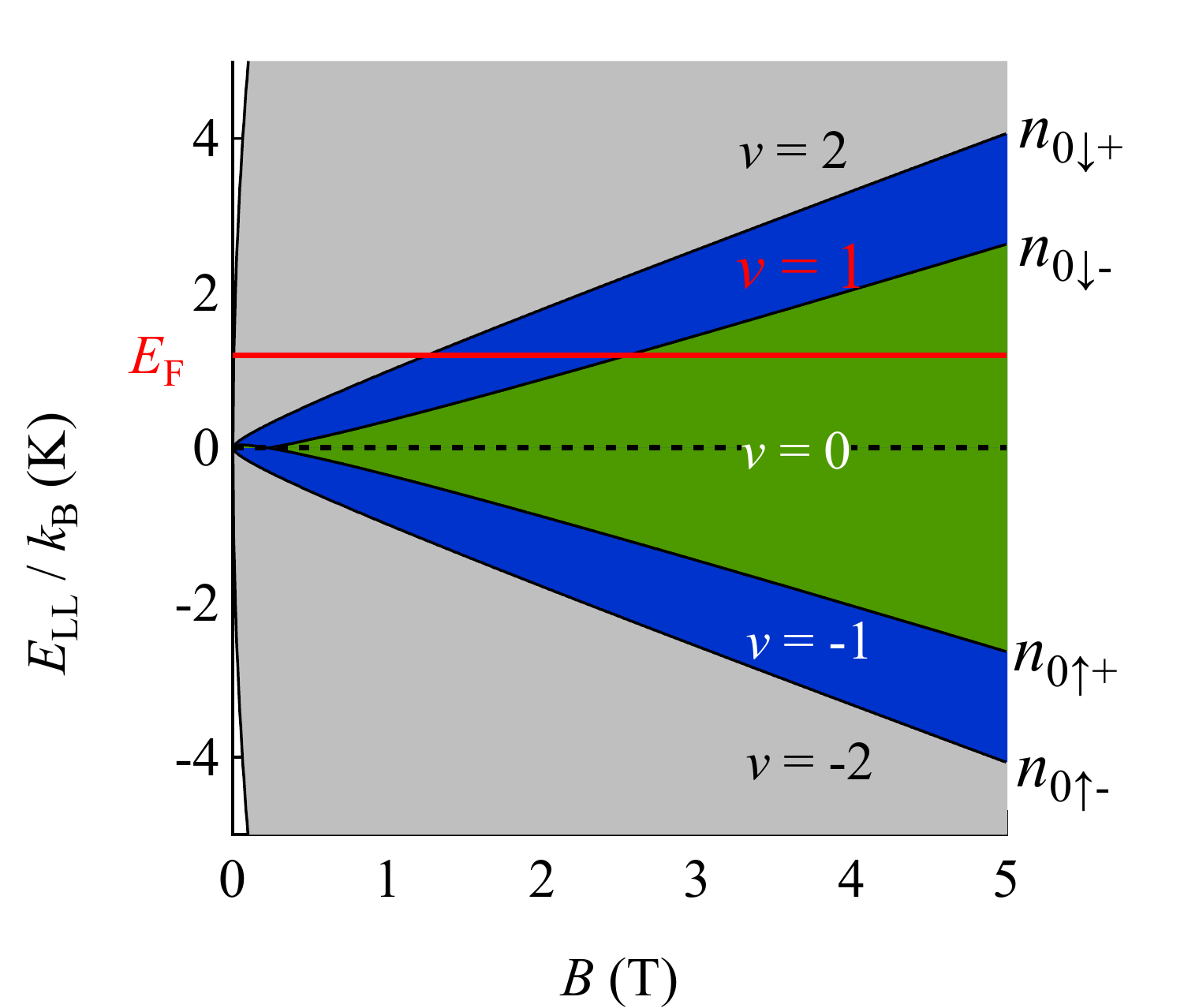}
  \caption{
    \label{fig:2}
    (color online) $n=0$ Landau level structure. $n_0$ denotes the $n=0$ Landau level, and the subscripts $\uparrow$ and $\downarrow$ denote spin splitting, and $+$ and $-$ denote valley splitting. 
  }
\end{figure}

Figure 3 shows the magnetic field dependence of resistivity $\rho_{xx}$ and Hall resistivity $-\rho_{xy}$ for the three samples under a pressure of approximately 1.7 GPa. 
Sample 1 was measured at 0.1 K and the other two were measured at 0.5 K. 
We notice SdH oscillation in $\rho_{xx}$ for all samples at magnetic fields below 3 T. 
The obvious plateaus observed in $-\rho_{xy}$ at magnetic fields of approximately 1.5 T show that $\rho_{xx}$ minima are the hallmarks of the quantum Hall effect.  

The breakdown of the plateaus of Hall resistivity due to the high current intensity is one of the characteristic features of quantum Hall effect\cite{rf:41, rf:42, rf:43}.
We verified that the plateaus were blurred with increasing current intensity, as shown in Fig. 3(d).
Note the data of $-\rho_{xy}$ in the normal state are independent of the current intensity. 
In the quantum Hall state, Joule heat is locally generated near the current electrode at the sample edge. 
Nonequilibrium electron distribution generated at the sample edge gives rise to energy dissipation so that the breakdown of the quantum Hall effect occurs. 

We examine the quantum Hall state that occurred around 1.5 T from the magnetic field dependence of the Hall conductivity per layer (sheet Hall conductivity), $\sigma_{xy}^{\rm sheet}$, calculated as $\sigma_{xy}^{\rm sheet}= \rho_{xy}/(\rho_{xx}^2+\rho_{xy}^2) \cdot c$ as shown in Fig. 3.
A clear plateau of $\sigma_{xy}^{\rm sheet}=e^2/h$ around 1.5 T for the three samples indicates that the $\nu =1$ quantum Hall state is realized.

\begin{figure}[htbp]
  \includegraphics[width=1 \linewidth]{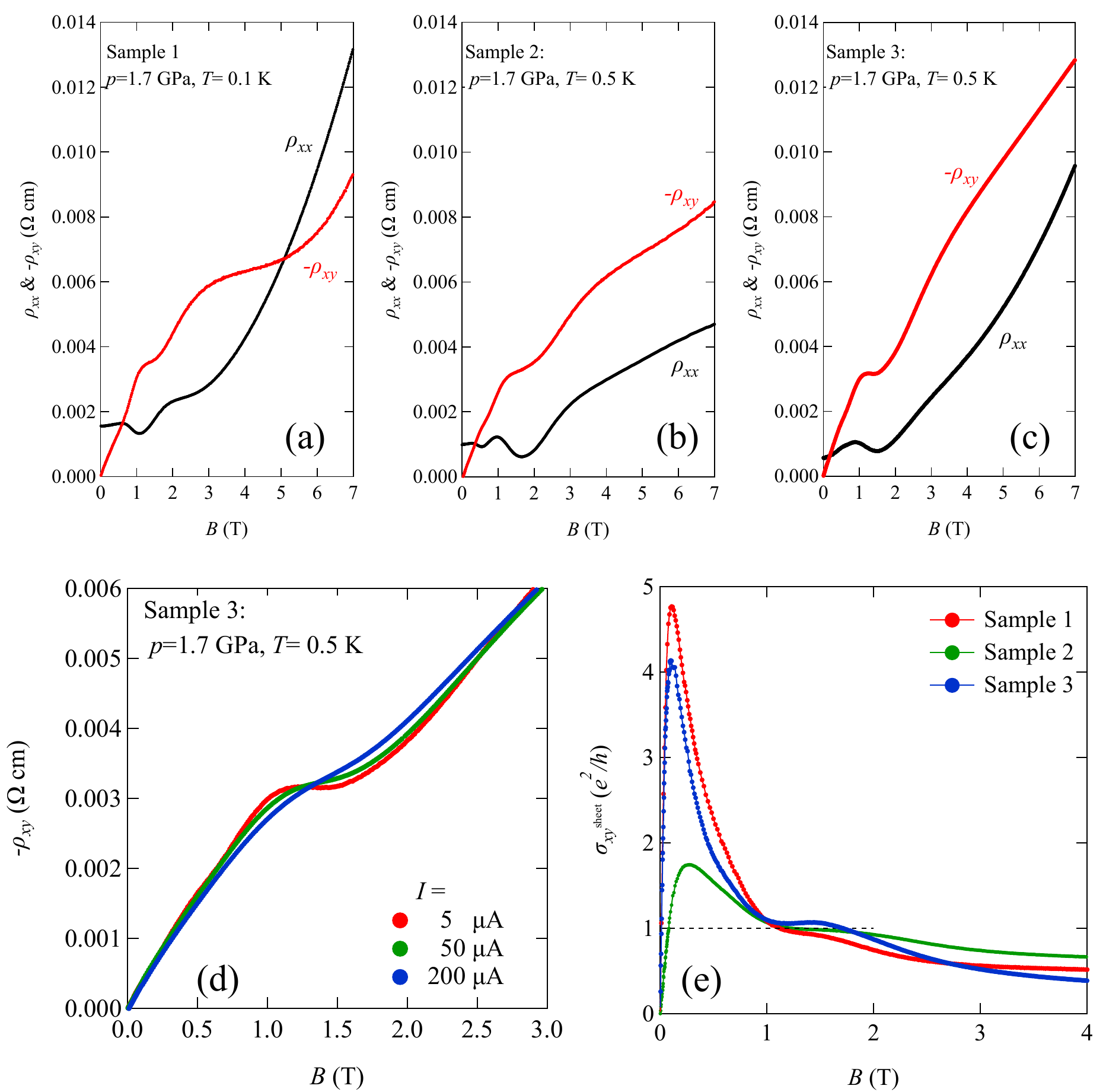}
  \caption{
    \label{fig:3}
     (color online) Magnetic field dependence of the resistivity and Hall resistivity of three $\alpha$-(BETS)$_2$I$_3$ samples [(a) Sample 1, (b) Sample 2 and (c) Sample 3] at low temperatures. 
     (d) The response of Hall resistivity to current intensity for sample 3. 
     The high current intensities blur the plateau of the Hall resistivity.
     (e) Magnetic field dependence of the sheet Hall conductivity $\sigma_{xy}^{\rm sheet}$ for three samples.
     Three samples show a $\sigma_{xy}^{\rm sheet}$ plateaus at $\sim e^2/h$. 
     This plateau indicates the $\nu = 1$ quantum Hall effect.
  }
\end{figure}

Finally, we briefly mention the inter-band effects of the magnetic field on the Hall conductivity in those systems.

A strong magnetic field has fascinated physicists the interesting phenomena such as the quantum Hall effect. 
On the other hand, a weak magnetic field also gives us characteristic phenomena. 
According to the theory of Fukuyama, the vector potential plays an important role in inter-band excitation in electronic systems with a vanishing or narrow energy gap \cite{rf:44}. 
The orbital movement of virtual electron-hole pairs gives rise to anomalous orbital diamagnetism and the Hall effect in a weak magnetic field. 
This effect is called the inter-band effect of the magnetic field and appears near the Dirac points.
To detect this effect on the Hall conductivity, we should control the chemical potential $\mu$. 
However, the conventional field-effect-transistor method has difficulty controlling $\mu$.
Hence, we have successfully detected inter-band effects of the magnetic field on the Hall conductivity by varying $\mu$ of $\alpha$-(BEDT-TTF)$_2$I$_3$ as follows\cite{rf:38}.

We return to Fig. 1(b) and focus on the dip structure that appeared in $R_H$ of $\alpha$-(BEDT-TTF)$_2$I$_3$ (S2) and the three $\alpha$-(BETS)$_2$I$_3$ samples.
In Fig. 1(b), the absolute value of $R_H$ is plotted. 
The dips indicate a change in the polarity of $R_H$.
The change in the polarity of $R_H$ for $\alpha$-(BEDT-TTF)$_2$I$_3$ was understood as follows.

This sample is classified as electron-doped or hole-doped depending on the ppm level dopant because of the instability of the I$_3$$^-$ anion.
In the hole-doped samples (S1), $R_H$ is positive over the whole temperature range. 
On the other hand, the polarity of $R_H$ in the electron-doped samples changes at low temperatures. 
In S2, for example, it is changed to 2 K. 

According to the first principal band calculation, the electron-hole symmetry at the system's low-energy region is imperfect\cite{rf:13}. 
In this situation, the chemical potential $\mu$ is no longer a constant and varies with the temperature. 
The polarity changes at a point where $\mu=0$.
Then, we obtained the experimental formula $\mu=E_F-0.024k_B T$, which quantitatively agrees with the theory\cite{rf:38, rf:45}.

Similarly, $\mu=0$ is expected at 52 K for $\alpha$-(BETS)$_2$I$_3$. 
However, the electron density estimated from $R_H$ at low temperatures 
is much higher than that of ppm-level dopants. 
Furthermore, the sample dependence of $R_H$ is weak.
There must be a reason other than I$_3$$^-$ anion instability 
for the Fermi energy to leave the Dirac points.

Assuming that the $\mu$ of $\alpha$-(BETS)$_2$I$_3$ is also linear with temperature, the experimental formula is obtained as $\mu = E_F - 0.027k_B T$ from $E_F/k_B=1.4$ K and a temperature of $\mu =0$. 
The agreement of two experimental formulas of $\mu$ indicates that the two crystals have the same electron-hole symmetry.

As a result, the temperature dependence of the Hall conductivity $\sigma_{xy}$ of $\alpha$-(BEDT-TTF)$_2$I$_3$ and $\alpha$-(BETS)$_2$I$_3$ shown in Fig. 4(a) at 0.01T can be plotted as a function of $\mu$ in Fig. 4(b). 
We see the peak structure in $\sigma_{xy}$ at the vicinity of $\mu =0$. 
It should be compared with the theoretical curves shown. 
Solid lines and dashed lines are the theoretical curves with and without the inter-band effects of the magnetic field, respectively.\cite{rf:45} 
The peak structure of $\sigma_{xy}$ originated from the inter-band effects of the magnetic field. 
The realistic theory indicates that the Hall conductivity without the inter-band effects of the magnetic field has no peak structure \cite{rf:45}.
The energy between two peaks is the damping energy that depends on the density of scattering centers in a crystal. 
Figure 4(b) indicates that the damping energy of $\alpha$-(BETS)$_2$I$_3$ is much higher than that of $\alpha$-(BEDT-TTF)$_2$I$_3$.
The intensity of the peak, on the other hand, depends on the damping energy and the tilt of the Dirac cones\cite{rf:45}. 
Note that the oversimplification of the empirical formula of $\mu$ causes the discrepancy between the experimental results and the theoretical curve at the positive $\mu$ side (low temperature).

The smooth change in the polarity of $\sigma_{xy}$ also shows that the two systems have intrinsic zero-gap structures. In a system with a finite energy gap, $\sigma_{xy}$ is changed in a stepwise manner\cite{rf:46}.

\begin{figure}[htbp]
  \includegraphics[width=1 \linewidth]{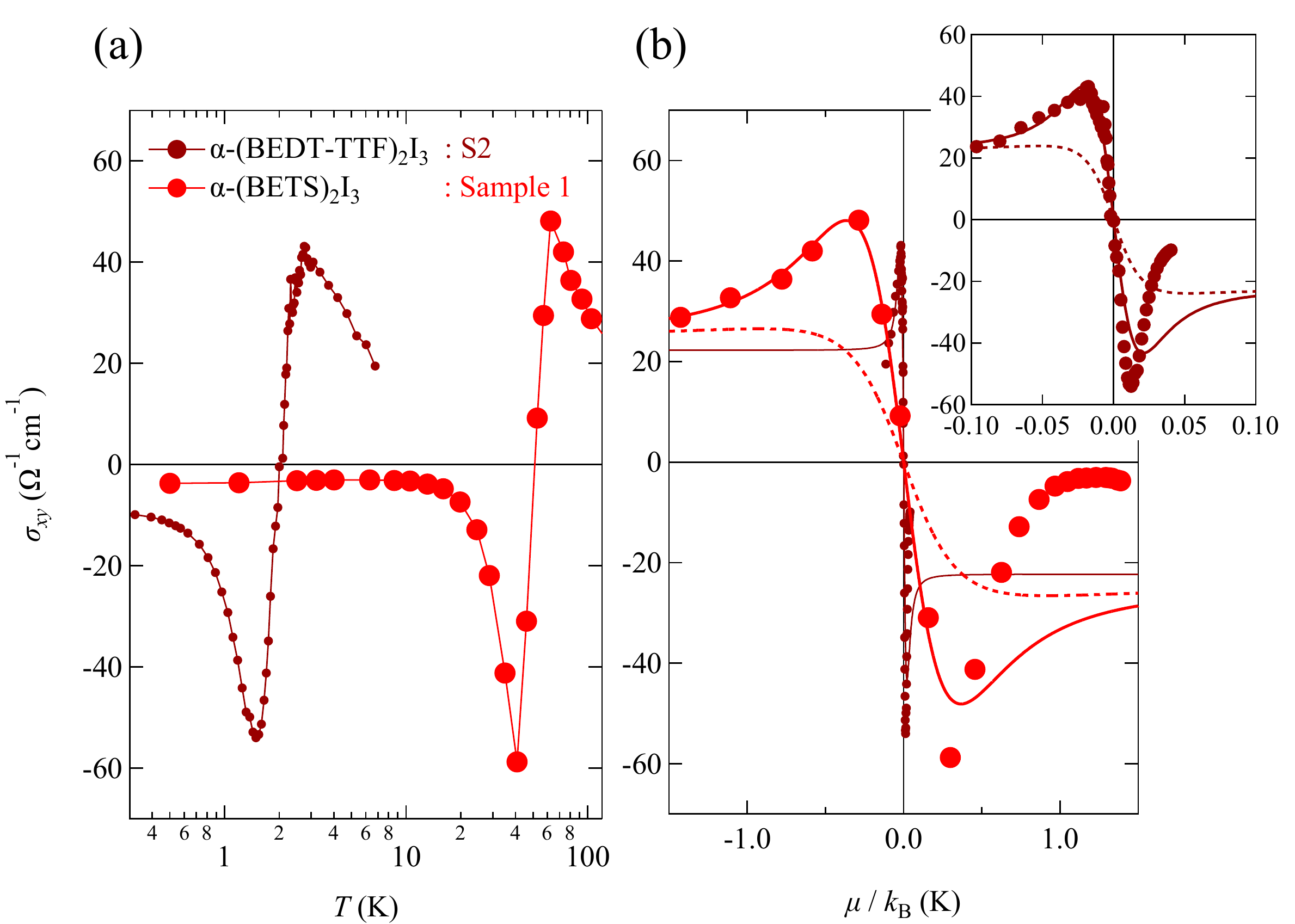}
  \caption{
    \label{fig:4}
     (color online) (a) Temperature dependence and (b) Chemical-potential dependence of $\sigma_{xy}$ for $\alpha$-(BEDT-TTF)$_2$I$_3$ (S2) and $\alpha$-(BETS)$_2$I$_3$ (Sample 1). 
     $\sigma_{xy}$ in $\alpha$-(BEDT-TTF)$_2$I$_3$ (S2) is enlarged in the inset.
     Solid lines and dashed lines are the theoretical curves with and without the inter-band effects of the magnetic field by Kobayashi {\it et al.}, respectively.\cite{rf:45} 
  }
\end{figure}

In conclusion, the Fermi energy of $\alpha$-(BETS)$_2$I$_3$ under the pressure of 1.7 GPa was approximately 1.4 K from the Dirac points. 
Thus the $\nu =1$ quantum Hall state was realized around 1.5 T at low temperatures.
It is rare for quantum Hall states to be realized in bulk crystals.
Furthermore, the temperature dependence of $\mu$ was clarified, and the inter-band effects of the magnetic field were successfully detected in $\sigma_{xy}$ when $\mu$ passes the Dirac point.

\acknowledgments{The research was supported by JSPS KAKENHI Grant Numbers 20K03870 and 22K03533.}


\end{document}